%% file: paper_1_final_TM.tex
\documentclass[12pt,a4paper,reqno]{article}
\usepackage[left=2cm,right=2cm,top=2cm,bottom=2cm]{geometry}

\usepackage{fullpage}

\usepackage{amssymb}
\usepackage{graphicx}
\usepackage{tikz,pgfplots,pgfplotstable}
\usepackage{color}
\usepackage{hyperref}             
\hypersetup{linktocpage}
\usepackage{versions}
\usepackage{frcursive}
\usepackage{mathrsfs}
\usepackage{mathtools,stmaryrd}

\usepackage[utf8]{inputenc}
\usepackage[english]{babel}
\usepackage{amsmath}
\usepackage{amsfonts}
\usepackage{amssymb}
\usepackage[titletoc,title]{appendix}
\usepackage{mathtools}
\usepackage{bbm}
\usepackage{bm}
\usepackage{subdepth}
\usepackage{enumerate}
\usepackage{mathrsfs}
\usepackage{graphicx}
\usepackage[outdir=./]{epstopdf}
\usepackage{esint}
\usepackage{dsfont}
\usepackage{amsthm}
\usepackage{cite}
\usepackage{url}
\usepackage{hyperref}
\usepackage{cleveref}
\usepackage{caption}
\hypersetup{%
        colorlinks=true,
        citecolor={black},
        linkcolor={black},
        urlcolor={black},}
\newcommand{\nm}{\noalign{\smallskip}}

\newcommand*{\RR}{\mathbb{R}}

\newcommand*{\xrarr}[2][]{\xrightarrow[#2]{#1}}

\newcommand*{\smin}{\setminus}

\newcommand*{\e}{\varepsilon}

\newcommand*{\pd}{\partial}

\renewcommand*{\a}{\alpha}

\newcommand{\G}{\Gamma}

\newcommand{\cqfd}{\hfill $\blacksquare$\\ \medskip}

\theoremstyle{plain}
\newtheorem{thm}{Theorem}[section]
\newtheorem{proposition}[thm]{Proposition}

\theoremstyle{definition}

\theoremstyle{remark}
\newtheorem{rmkx}[thm]{Remark}

\renewcommand{\o}{\omega}
\renewcommand{\l}{\lambda}
\renewcommand{\O}{\Omega}

\makeatletter
\def\blfootnote{\gdef\@thefnmark{}\@footnotetext}
\makeatother

\begin{document}
\title{Perturbation of the scattering resonances of an open cavity by small particles. Part I: The transverse magnetic polarization case}

\author{
Habib Ammari\thanks{\footnotesize Department of Mathematics,
ETH Z\"urich,
R\"amistrasse 101, CH-8092 Z\"urich, Switzerland (habib.ammari@math.ethz.ch, alexander.dabrowski@sam.math.ethz.ch).} \and
Alexander Dabrowski\footnotemark[1] \and Brian Fitzpatrick\thanks{\footnotesize
ESAT - STADIUS,
Stadius Centre for Dynamical Systems,
Signal Processing and Data Analytics,
Kasteelpark Arenberg 10 - box 2446,
3001 Leuven,
Belgium (bfitzpat@esat.kuleuven.be).}
\and Pierre Millien\thanks{\footnotesize  Institut Langevin,  1 Rue Jussieu, 75005 Paris, France (pierre.millien@espci.fr).} }

\date{}

\maketitle

\begin{abstract}
This paper aims at providing a small-volume expansion framework for the scattering resonances of an open cavity perturbed by small particles. The induced shift of the scattering frequencies by the small particles is derived without neglecting the radiation effect. The formula holds for arbitrary-shaped particles. It shows a strong enhancement in the frequency shift in the case of plasmonic particles. The formula is used to image small particles located near the boundary of an open resonator which admits whispering-gallery modes. Numerical examples of interest for applications are presented.
\end{abstract}

\def\keywords2{\vspace{.5em}{\textbf{Mathematics Subject Classification
(MSC2000).}~\,\relax}}
\def\endkeywords2{\par}
\keywords2{35R30, 35C20.}

\def\keywords{\vspace{.5em}{\textbf{Keywords.}~\,\relax}}
\def\endkeywords{\par}
\keywords{Open cavity, shift of scattering resonances, whispering-gallery modes, bio-sensing, plasmonic nanoparticles.}

\section{Introduction}

The influence of a small particle on a cavity mode plays an important role in fields such as optical sensing, cavity quantum electrodynamics, and cavity optomechanics \cite{haroche1,koenderink1,agallery1}. In this paper, we consider the transverse magnetic polarization case and provide a formal derivation of the perturbations
of scattering resonances of an open cavity due to a small-volume particle without neglecting the radiation effect. Note that the radiation effect has been omitted in the physics literature (see, for instance, \cite{WG1}). Indeed, the Bethe-Schwinger closed cavity perturbation formula  \cite{volkov, bethe} has been widely employed for radiating cavities.
The small-volume asymptotic formula in this paper generalizes to the open cavity case those derived in \cite{lee,khelifi, moskow,volkov}. It is valid for arbitrary-shaped particles. It shows that the perturbations of the scattering resonances can be expressed in terms of the polarization tensor of the small particle. Two cases are considered: the one-dimensional case and the multi-dimensional case. Its applicability to the perturbations of whispering-gallery modes by external arbitrary-shaped particles is also discussed. Finally, we characterize the effect that a plasmonic nanoparticle, of arbitrary geometry and which is bound to the surface of the cavity, has on the whispering-gallery modes of the cavity. Since the shift of the scattering frequencies is proportional to the polarization of the plasmonic nanoparticles \cite{plasmonic1,AmmariMillien2018, plasmonic2,plasmonic3}, which blows-up at the plasmonic resonances, the effect of a plasmonic particle on the cavity modes can be significant.

For the analysis of the transverse electric case we refer the reader to \cite{paper2}. Note that in the one-dimensional case, the scattering resonances are simple while in the multi-dimensional case, they can be degenerate or even exceptional. For the analysis of exceptional points, we again refer the reader to \cite{paper2}. The analysis of such a challenging problem is much simpler in the transverse electric case than in the transverse magnetic one. The reader is also referred to \cite{heider, HeiderBerebicezKohnWeinstein, santosa} for small amplitude sensitivity analyses of the scattering resonances. Numerical computation of resonances has been addressed, for instance,  in \cite{gop,kim,lin1,lin2,osting,majda}. 

The paper is organized as follows. In Section \ref{sec-1}, using the method of matched asymptotic expansions, we derive the leading-order term in the shifts of scattering resonances of a one-dimensional open cavity and characterize the effect of radiation. Section \ref{sec-2} generalizes the method to the multi-dimensional case.  In Section \ref{sec-3}, we consider the perturbation of whispering-gallery modes by small particles.  The formula obtained for the shifting of the frequencies shows a strong  enhancement in the frequency shift in the case of plasmonic particles, which allows for their recognition in spite of their small size. The splitting  of scattering frequencies of the open cavity of multiplicity greater than one due to small particles is also discussed.  In Section \ref{sec-4}, we present some numerical examples to illustrate the accuracy of the formulas derived in this paper and their use in the sensing of small particles. The paper ends with some concluding remarks.

\section{One dimensional case} \label{sec-1}

\begin{figure}[h!]
\def\svgwidth{0.9\linewidth}
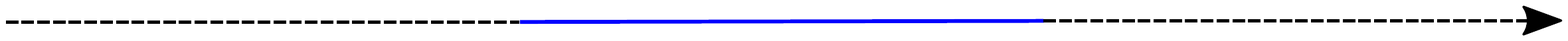
\caption{\footnotesize{One dimensional cavity.}}
\label{figadd1}
\end{figure}

We first consider a one dimensional cavity. We let the magnetic permeability $\mu_\delta$ be $\mu_m$ in $(a,b) \smin (-\delta/2,\delta/2)$ and  $\mu_c$ in $ (-\delta/2,\delta/2)$ and the electric permittivity $\e_\delta $ be $\e_m$ in $(a,b) \smin (-\delta/2,\delta/2)$ and  $\e_c$ in $ (-\delta/2,\delta/2)$, see Figure \ref{figadd1}. Here, $0<\delta < 1/2$ and $\mu_m,\mu_c, \e_m,$ and $\e_c$ are positive constants. 

Let $\omega_0$ be a scattering resonance of the unperturbed cavity and let $u_0$ denote the corresponding eigenfunction, that is, 

\begin{equation*} \begin{cases}
\partial_x \left( (1/\varepsilon_m ) \partial_x u_0\right) + \o_0^2 \mu_m u_0 = 0 & \text{in } (a,b), \\
(1/\e_m) \partial_x u_0 + i \o_0 u_0 = 0 & \text{at } a,\\
(1/\varepsilon_m) \partial_x u_0 - i \o_\delta u_0 = 0 & \text{at } b,\\
\int_{-1/2}^{1/2} |u_0|^2 \, dx=1.
\end{cases}
 \end{equation*}

We now consider the perturbed problem: for $\delta$ small, we seek a solution $u_\delta$, for
which $\omega_\delta \rightarrow \omega_0$ as $\delta \rightarrow 0$ of the following  equation:
\begin{equation} \label{sc1} \begin{cases}
\partial_x \left( (1/\varepsilon_\delta) \partial_x u_\delta\right) + \o_\delta^2 \mu_\delta u_\delta = 0 & \text{in } (a,b), \\
(1/\e_m) \partial_x u_\delta + i \o_\delta u_\delta = 0 & \text{at } a,\\
(1/\varepsilon_m) \partial_x u_\delta - i \o_\delta u_\delta = 0 & \text{at } b,\\
\int_{-1/2}^{1/2} |u_\delta|^2 \, dx=1.
\end{cases} \end{equation}
\begin{rmkx}
The above one-dimensional scattering resonance problems  govern scattering resonances of slab-type structures. They are a consequence of Maxwell's equations, under the assumption of time-harmonic solutions. They correspond to the transverse magnetic polarization; see \cite{HeiderBerebicezKohnWeinstein}. The scattering resonances $\omega_0$ and $\omega_\delta$ lie in the lower-half of the complex plane. The eigenfunctions $u_0$ and $u_\delta$ satisfy the outgoing radiation conditions at $a$ and $b$ and, consequently, grow exponentially at large distances from the cavity. To give a physical interpretation of scattering resonances, we must go to the time domain, see, for instance, \cite{gop, HeiderBerebicezKohnWeinstein}.    

%Please note that is this paper, we assume that the speed of the wave is $1$ outside the cavity. Of course,  it is not a restrictive assumption, and if one wants to use realistic values for $\varepsilon$ and $\mu$ one can replace $\omega$ by $\omega/c_0$ where $c_0$ is the real speed of light in vacuum. The analysis remains valid and all formulas derived in this paper can be used.
\end{rmkx}

\begin{proposition} As $\delta \rightarrow 0$, we have
$$\omega_\delta = \omega_0 + \delta \omega_1 +O(\delta^2),$$ where
\begin{equation}
\label{eq:formula1d}
\o_1 = \dfrac{\a (\partial_x u_0(0))^2 + \o_0^2 \varepsilon_m (\mu_c - \mu_m) (u_0(0))^2}{2 \o_0 \mu_m \e_m \int_{-1/2}^{1/2} u_0^2\, dx + i \e_m ((u_0(a))^2 + (u_0(b))^2)}.\end{equation}
The polarization $\alpha$ is defined by
\begin{equation} \label{alphad1}
\alpha= \left( \frac{\varepsilon_m}{\varepsilon_c} -1\right)\partial_x v^{(1)} (\frac{1}{2}) \big|_{-}, 
\end{equation} and $v^{(1)}$ is the unique solution (up to a constant) of the auxiliary  differential equation:
\begin{align*}
\left\{\begin{aligned}
&\partial_x ({1}/{\tilde{\varepsilon}}) \partial_x v^{(1)} =0, \\ 
&v^{(1)}(\xi) \sim \xi \quad \vert \xi\vert \rightarrow + \infty, 
\end{aligned}\right. 
\end{align*}
{with} $\tilde{\varepsilon}=\varepsilon_c \chi_{(-1/2, 1/2)} + \varepsilon_m \chi_{\mathbb{R}\setminus(-1/2, 1/2) }$. Here, $|_{-}$ indicates the limit at $(1/2)^-$ and $\chi_I$ denotes the characteristic function of the set $I$. 
\end{proposition}
\begin{rmkx}
Note that the polarization $\alpha$ can be computed explicitly. It is given by $\alpha = 1- (\e_c/\e_m)$. 
\end{rmkx}

\proof
Using the method of matched asymptotic expansions for $\delta$ small, see \cite{khelifi}, we construct asymptotic expansions of $\omega_\delta$ and $u_\delta$.

To reveal the nature of the perturbations in $u_\delta$, we introduce the local
variable $\xi = x/\delta$ and set $e_\delta(\xi)  = u_\delta(x)$. We expect that $u_\delta(x)$
will differ appreciably from $u_0(x)$ for $x$ near $0$, but it will differ little from $u_0(x)$  for $x$ far from $0$. Therefore, in the spirit of matched asymptotic expansions, we shall represent $u_\delta$ by two different expansions, an inner expansion for $x$ near $0$, and an outer
expansion for $x$ far from $0$. We write
the outer and inner expansions:
$$u_\delta(x) = u_0(x) + \delta u_1(x) + \dots \quad  \mbox{for } |x| \gg \delta , $$ and  $$u_\delta(x) = e_0(\xi) + \delta e_1(\xi) + \dots \quad \mbox{for } |x| = O(\delta).$$
The asymptotic expansion of $\omega_\delta$ must begin with $\omega_0$, so we write
$$ \omega_\delta = \omega_0 + \delta \omega_1 + \ldots . $$
In order to determine the functions $u_i(x)$ and $e_i(\xi)$, we
have to equate the inner and the outer expansions in some “overlap” domain within which
the stretched variable $\xi$ is large and $x$ is small. In this domain the matching conditions are:
$$u_0(x) + \delta u_1(x) + \dots \sim e_0(\xi) + \delta e_1(\xi) + \dots.$$
Now, if we substitute the
inner expansion into (\ref{sc1}) and formally equate coefficients
of  $\delta^{-2}$ and $\delta^{-1}$, then we obtain
 $$\pd_\xi ( (1/\tilde \varepsilon) \pd_\xi e_0) = 0, $$ and $$\pd_\xi ( (1/\tilde \varepsilon) \pd_\xi e_1) = 0,$$ where the stretched coefficient $\tilde \varepsilon $ is equal to $ \varepsilon_c$ in $(-1/2,1/2)$ and to $\varepsilon_m$ in $(-\infty, -1/2) \cup (1/2, +\infty)$.
From the first matching
condition, it follows that $e_0(\xi) = u_0(0)$ for all $\xi$.
Similarly, we have \begin{equation} \label{secondmatchh} 
e_1(\xi) \sim  \xi \partial_x u_0(0) \quad \mbox{as } |\xi| \to + \infty.\end{equation}

Let $v^{(1)}(\xi) $ be such that
$$\begin{cases}
\pd_\xi ( (1/\tilde \varepsilon(\xi)) \pd_\xi v^{(1)}(\xi)) = 0, \\
v^{(1)}(\xi) \sim \xi  \quad \text{as } |\xi| \to + \infty.
\end{cases}$$
Let $G(\xi) = |\xi |/2$ be the free space Green function,
$$\pd^2_{\xi} G(\xi-\xi') = \delta_0(\xi - \xi').$$
Since $$\pd^2_{\xi} v^{(1)}(\xi) = (1 - (\varepsilon_m / \varepsilon_c)) \partial_\xi v^{(1)}(-1/2) |_+ + ((\varepsilon_m / \varepsilon_c) -1)  \partial_\xi  v^{(1)}(1/2) |_-,$$
we have
$$v^{(1)}(\xi) = \xi + (1 - (\varepsilon_m / \varepsilon_c) ) \partial_\xi v^{(1)}(-1/2)|_+ G(\xi + 1/2) + ((\varepsilon_m/ \varepsilon_c) -1) \partial_\xi v^{(1)}(1/2) |_- G(\xi - 1/2),$$ where the subscripts $+$ and $-$ indicate the limits at $(1/2)^-$ and $(1/2)^+$, respectively. 
Moreover, $$\int_{-1/2}^{1/2} \partial^2_\xi v^{(1)} \, d\xi= 0,$$ yields $$ \partial_\xi  v^{(1)}(-1/2) |_+
=  \partial_\xi  v^{(1)}(1/2) |_-.$$  Hence,
$$v^{(1)}(\xi) = \xi + ((\varepsilon_m/ \varepsilon_c) - 1)  \partial_\xi  v^{(1)}(1/2) |_- G(\xi + 1/2) - 
((\varepsilon_m/ \varepsilon_c) -1)  \partial_\xi  v^{(1)}(1/2)|_- G(\xi - 1/2) .$$
On the other hand,  $$G(\xi - 1/2) \sim |\xi| - \xi /(2 |\xi|) + \dots ,$$ and $$G(\xi + 1/2) \sim |\xi| + \xi / (2 |\xi|) + \dots \quad \mbox{as } |\xi| \to + \infty.$$
Therefore,
$$v^{(1)}(\xi) \sim \xi - ((\varepsilon_m/\varepsilon_c) -1)  \partial_\xi  v^{(1)}(1/2)|_- \; \xi/ |\xi| + \dots .$$
The second matching condition  (\ref{secondmatchh})  yields $$u_1(x) \sim  \bigg( - \partial_x u_0(0)  ((\varepsilon_m/ \varepsilon_c) -1)  \partial_\xi  v^{(1)}(1/2) |_- \bigg)  \xi / |\xi| \quad \mbox{for } x  \mbox{ near } 0.$$

Assume first that $\mu_m = \mu_c$. To find the first correction $\omega_1$, we multiply $$\partial_x ( (1/\varepsilon_m) \partial_x u_1) + \o_0^2 \mu_m u_1 = - 2 \o_1 \o_0 \mu_m u_0$$ by ${u}_0$ and integrate over $(a, - \rho/2)$ and $(\rho/2,b)$ for $\rho$ small enough.  Upon using the radiation condition and  Green's theorem, we obtain as $\rho$ goes to zero, 
$$i \o_1 ((u_0(a))^2 + (u_0(b))^2) -\frac{1}{\e_m} \a (\partial_x u_0(0))^2 = -2 \o_1 \o_0 \mu_m \int_{-1/2}^{1/2} u_0^2 \, dx,$$
where the polarization $\alpha$ is given by \begin{equation}
\label{defalpha}  \a = ( (\varepsilon_m / \varepsilon_c) -1)  \partial_\xi v^{(1)}(1/2)|_-  = 1 - \frac{\e_c}{\e_m}. \end{equation} Therefore, we arrive at
\begin{equation}\label{eq:onedimensionalpartial}
 \o_1 = \dfrac{\a (\partial_x u_0(0))^2}{2 \o_0 \mu_m \e_m \int_{-1/2}^{1/2} u_0^2 \, dx + i \e_m ((u_0(a))^2 + (u_0(b))^2)}.\end{equation}
The term $i \e_m ((u_0(a))^2 + (u_0(b))^2)$ accounts for the effect of radiation on the shift of the scattering resonance $\omega_0$.

Now, if $\mu_c \neq \mu_m$, then we need to compute the second-order corrector $e_2$.
We have
$$\pd_\xi ( (1/\tilde \varepsilon) \pd_\xi e_2) + \o_0^2 \tilde \mu  e_0 = 0, $$ and $$e_2(\xi) \sim \xi^2 \partial^2_x u_0(0)/2 \quad \mbox{as } |\xi| \to + \infty.$$ Here, the stretched coefficient $\tilde \mu $ is equal to $ \mu_c$ in $(-1/2,1/2)$ and to $\mu_m$ in $(-\infty, -1/2) \cup (1/2, + \infty)$.

From the equation satisfied by $u_0$, we obtain
$$\partial^2_x u_0 (0) = -\o_0^2 \mu_m \varepsilon_m  u_0(0).$$
Recall that $e_0(\xi) = u_0(0)  $ and let $v^{(2)}$ be such that
$$\begin{cases}
\pd_\xi ( (1/\tilde \varepsilon(\xi)) \pd_\xi v^{(2)}(\xi)) = (1/(\varepsilon_m \mu_m)) \tilde \mu (\xi), \\
v^{(2)}(\xi) \sim \xi^2/2  \quad \text{as } |\xi| \to + \infty.
\end{cases}$$
It is easy to see that $\pd_\xi ( (1/\tilde \varepsilon(\xi)) \pd_\xi (v^{(2)}(\xi) - \xi^2/2)) $ is $ (1/\varepsilon_m) ((\mu_c/ \mu_m) -1) $ for $\xi \in (-1/2,1/2) $ and is $0$ for $|\xi| > 1/2.$
Therefore, $$v^{(2)}(\xi) - \xi^2/2 \sim ( (\mu_c / \mu_m) -1) |\xi| \quad \mbox{as } |\xi|\to + \infty.$$
Then $$u_1(x) \sim  \partial_x u_0(0) ( \xi - ((\varepsilon_m / \varepsilon_c) -1)  \partial_\xi v^{(1)}(1/2) \xi / |\xi| + \dots) + \partial^2_x u_0(0) ( (\mu_c/\mu_m) -1) |\xi| + \dots,$$
and so
$$i \o_1 ((u_0(a))^2 + (u_0(b))^2) - \frac{1}{\e_m} \a (\partial_x u_0(0))^2 + \frac{1}{\e_m} \partial^2_x u_0(0) ((\mu_c/\mu_m) -1) {u}_0(0) = - 2 \o_1 \o_0 \mu_m \int_{-1/2}^{1/2} u_0^2 \, dx, $$ which yields the  result. \cqfd

\section{Multi-dimensional case} \label{sec-2}

In this section,  we generalize (\ref{eq:formula1d}) to the multi-dimensional case. In dimension two, the obtained formula corresponds, as in the one-dimensional case, to an open cavity with the transverse magnetic polarization \cite{santosa}.  We use the same notation as in Section \ref{sec-2}.

\begin{figure}[h!]
\def\svgwidth{0.9\linewidth}
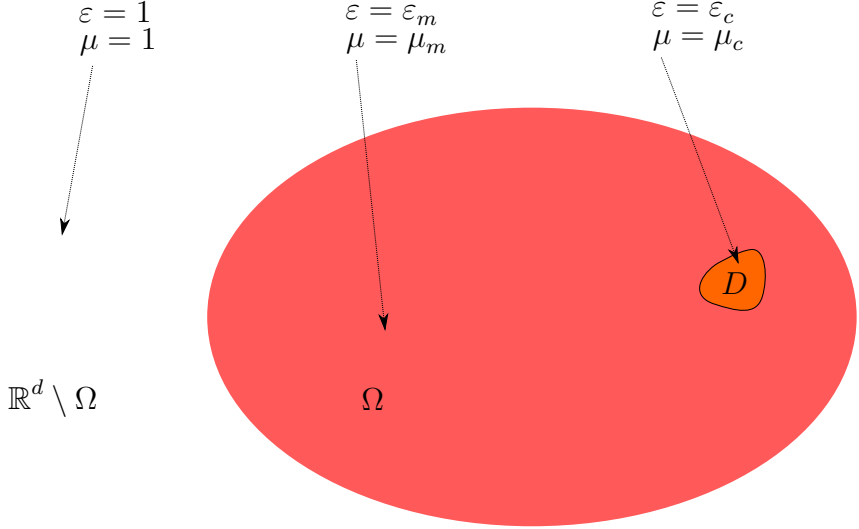
\caption{\footnotesize{Multi-dimensional cavity.}}
\label{figadd2}
\end{figure}

Let $\Omega$ be a bounded domain in $\mathbb{R}^d$ for $d=2,3,$ with smooth boundary $\partial \Omega$, see Figure \ref{figadd2}. Let $\o_0$ be a simple eigenvalue of the unperturbed open cavity. 
Then there exists a non trivial solution $u_0$ to the equation:

\begin{align} \label{eq:defomega0}
\left\{\begin{aligned}
&\nabla \cdot \left( ({1}/{\varepsilon}) \nabla u\right) + \omega_0^2 \mu u =0 \quad \mbox{in } \mathbb{R}^d,  \\
&\int_\Omega |u|^2 \, dx =1,\\
&u \text{\ satisfies the outgoing radiation condition},  
\end{aligned}\right.
\end{align}
where $\mu=1 + (\mu_m-1) \chi_\Omega $ and $\varepsilon= 1 + (\varepsilon_m-1) \chi_\Omega$. Here, $\chi_\Omega$ denotes the characteristic function of the domain $\Omega$. We refer to \cite{gop} for a precise statement of the outgoing radiation condition. 

For simplicity, we assume that $\Omega$ is the ball of radius $R$ centered at the origin, and introduce the capacity operator $T_\o$, which is given by \cite{nedelec}
$$
T_\o:  \phi = \left\{ \begin{array}{l} \displaystyle \sum_{m\in \mathbb{Z}} \phi_m e^{im\theta} \\
\nm \displaystyle
\sum_{m=0}^{+\infty} \sum_{l=-m}^m \phi_m^l Y^l_m \end{array}  \right. \mapsto
\left\{ \begin{array}{l} \displaystyle \sum_{m\in \mathbb{Z}} z_m(\omega,R) \phi_m e^{im\theta},\\
\nm \displaystyle
\sum_{m=0}^{+\infty}  z_m(\omega,R) \sum_{l=-m}^m \phi_m^l Y^l_m, \end{array} \right.
$$
where
$$
z_m(\omega, R)= \left\{ \begin{array}{l} \displaystyle \frac{\omega (H^{(1)}_m)^\prime (\omega R)}{H^{(1)}_m (\omega R)}  \quad \mbox{if } d=2,\\
\nm \displaystyle \frac{\omega (h^{(1)}_m)^\prime (\omega R)}{h^{(1)}_m (\omega R)}  \quad \mbox{if } d=3. \end{array} \right.
$$
Here, $\theta$ is the angular variable, $Y_m^l$ is a spherical harmonic, and $H_m^{(1)}$ (respectively, $h_m^{(1)}$) is the Hankel function of integer order (respectively, half-integer order).
This explicit version of the capacity operator will be used in Section \ref{sec-4} to test the validity of our formula. Then, \eqref{eq:defomega0} is equivalent to

\begin{equation}\label{eq:defomega0T}\begin{cases}
(1/\varepsilon_m) \Delta u_0 + \o_0^2\mu u_0 = 0 \quad \mbox{in } \O, \\
(1/\varepsilon_m)\dfrac{\pd u_0}{\pd \nu} = T_{\o_0} [u_0] \quad  \mbox{on }  \pd \O, \\
\int_\O |u_0|^2 = 1, 
\end{cases}\end{equation}
where $\nu$ denotes the normal to $\partial \Omega$. As in the one-dimensional case, the scattering resonances lie in the lower-half of the complex plane and the associated eigenfunctions grow exponentially at large distances from the cavity since they satisfy the outgoing radiation condition.  
We also remark that since on one hand, $z_{-m}(\omega, R) = z_m(\omega, R)$ for all $m\in \mathbb{Z}$, and on the other hand, $Y_m^{-l}= (-1)^l \overline{Y}^{l}_m$,  we have
\begin{equation} \label{addtc} \int_{\pd \O} T_\o [f] {g}\, d\sigma  = \int_{\pd \O} f {T_\o[g]} \, d\sigma \quad \mbox{for all } f, g \in  H^{1/2} (\pd \O),\end{equation} for $d=2,3$, where $H^s(\partial \Omega)$ is the standard Sobolev space of order $s$.

Let $D \Subset \Omega$ be a small particle of the form $D= z + \delta B$, where $\delta$ is its characteristic size, $z$ its location, and $B$ is a smooth bounded domain containing the origin. Denote respectively by $\e_c$ and $\mu_c$ the electric permittivity and the magnetic permeability of the particle $D$. The eigenvalue problem is to find $\o_\delta$ such that there is a non-trivial couple $(\o_\delta,u_\delta)$ satisfying
$$\begin{cases}
(1/\varepsilon_m) \Delta u_\delta + \o_\delta^2\mu_m u_\delta = 0 & \mbox{in }  \O \smin \bar D, \\
(1/\varepsilon_c) \Delta u_\delta + \o_\delta^2\mu_c u_\delta = 0 & \mbox{in }  D, \\
(1/\varepsilon_m) \dfrac{\pd u_\delta}{\pd \nu} \big|_{+} = (1/\varepsilon_c) \dfrac{\pd u_\delta}{\pd \nu} \big|_{-}  & \mbox{on }  \pd D, \\
(1/\varepsilon_m) \dfrac{\pd u_\delta}{\pd \nu} = T_{\o_\delta} [u_\delta] & \mbox{on }  \pd \O,
\end{cases}$$
where the subscripts $+$ and $-$ indicate the limits from outside and inside $D$, respectively. 

\begin{proposition} As $\delta\rightarrow 0$, we have $$\omega_\delta = \omega_0 + \delta^d \omega_1 +O(\delta^{d+1}),$$ where
\begin{equation}
\label{finaleqeq2} \displaystyle \omega_1 = \dfrac{ M(\varepsilon_m / \varepsilon_c, B) \nabla u_0(z) \cdot \nabla {u}_0 (z) +  \o_0^2 |B|  \varepsilon_m (\mu_c -\mu_m) (u_0(z))^2}{2\o_0 \mu_m \e_m \int_\Omega u_0^2 \, dx +  \, \e_m  \int_{\pd \O} \pd_\o T_\o  |_{\o = \o_0} [u_0] {u}_0\, d\sigma } ,\end{equation} where $M$ is the polarization tensor  associated with the domain $B$ and the contrast $\varepsilon_m/ \varepsilon_c$ defined by (\ref{defmp1}) with $v^{(1)}$ being given by (\ref{pbv}). Note that $M$ has the same form as $\alpha$ defined in (\ref{alphad1}). 
\end{proposition}

\proof Assume, for now, that $\mu_c=\mu_m$.
Let $\l_0 = \o_0^2, \l_\delta = \o_\delta^2$.
We expand $$\omega_\delta = \omega_0 +\delta^d \omega_1 +\ldots \quad \mbox{and } \quad \l_\delta = \l_0 + \delta^d \l_1 + \dots .$$
Let the outer expansion of $u_\delta$ be $$u_\delta(y) = u_0(y) + \delta^d u_1(y) + \dots,$$ and the inner one,
$e_\delta(\xi) = u_\delta ((x-z)/\delta)$, be $$e_\delta (\xi) = e_0(\xi) + \delta e_1(\xi) + \dots .$$
Therefore, we have $$T_{\o_\delta} \simeq T_{\o_0 + \delta^d \o_1} \simeq T_{\o_0} + \delta^d \o_1 \pd_\o  T_\o |_{\o_0} + \dots .$$
Moreover, we obtain
\begin{equation} \label{et1} \begin{cases}
( (1/\varepsilon_m) \Delta + \l_0 \mu_m ) u_1(y)  = - \l_1 \mu_m u_0(y) & \mbox{for } |y-z| \gg O(\delta) , \\
(1/\varepsilon_m) \dfrac{\pd u_1}{\pd \nu} = T_{\o_0} [u_1] + \o_1  \pd_\o  T_\o|_{\o = \o_0} [u_0] &  \mbox{on }  \pd \O,
\end{cases} \end{equation}
and
$$\begin{cases}
 \Delta_\xi e_j = 0 &  \mbox{in } \RR^d \smin \bar B,\\
 \Delta_\xi e_j = 0 &  \mbox{in }  B, \\
 \dfrac{\pd e_j}{\pd \nu}|_+ =  (\e_m/\varepsilon_c) \dfrac{\pd e_j}{\pd \nu} |_- &  \mbox{on } \pd B,
\end{cases}$$
for $j=1,2.$
Imposing the matching conditions $$u_0(y) + \delta^d u_1(y) + \dots \sim e_0(\xi) + \delta e_1(\xi) + \dots \quad \mbox{as } |\xi| \to + \infty , $$ and $y \to z$, we arrive at
$e_0(\xi) \to u_0(z)$ and $e_1(\xi) \sim \nabla u_0(z) \cdot \xi$.
So, we have $e_0(\xi) = u_0(z)$ for every $\xi$ and $e_1(\xi) = \nabla u_0(z) \cdot v^{(1)}(\xi)$, where $v^{(1)}$ is such that (see \cite{khelifi})
\begin{equation} \label{pbv} \begin{cases}
\Delta_\xi v^{(1)} = 0 &  \mbox{in } \RR^d \smin \bar B,\\
 \Delta_\xi v^{(1)} = 0 &  \mbox{in }  B, \\
\dfrac{\pd v^{(1)}}{\pd \nu}|_+ = (\e_m/\varepsilon_c) \dfrac{\pd v^{(1)}}{\pd \nu} |_- &  \mbox{on } \pd B,\\
v^{(1)}(\xi) \sim \xi & \mbox{as } |\xi| \rightarrow +\infty.
\end{cases} \end{equation}
 Let $\Gamma$ be the fundamental solution of the Laplacian in $\mathbb{R}^d$. Let $M(\varepsilon_m/ \varepsilon_c, B)$ be the polarization tensor associated with the domain $B$ and the contrast $\varepsilon_m/ \varepsilon_c$ given by \cite{kang_book}
 \begin{equation} \label{defmp1}
 M(\varepsilon_m/ \varepsilon_c, B) =  (\frac{\varepsilon_m}{\varepsilon_c} -1) \int_{\partial B} \frac{\partial v^{(1)}}{\partial \nu} \big|_{-} (\xi) \xi \, d\sigma(\xi).
 \end{equation}
Then, by the same arguments as in \cite[Section 4.1]{khelifi},
it follows that \begin{equation} \label{et2} \ u_1(y) \sim - M(\varepsilon_m/ \varepsilon_c, B) \nabla \G(y-z) \cdot \nabla u_0(z) \quad \mbox{as } y \rightarrow z. \end{equation}
Multiplying (\ref{et1}) by ${u}_0$ and integrating by parts over $\O \smin \bar B_\delta$, we obtain from (\ref{addtc}) that
$$\begin{array}{lll}
\displaystyle -\l_1 \mu_m \int_{\O \smin B_\rho} (u_0)^2  \, dx &=& \displaystyle \underbrace{\int_{\pd \O} \big( T_{\o_0} [u_1] {u}_0 - {T_{\o_0} [u_0]} u_1 \big) \, d\sigma}_{=0} +  \o_1 \int_{\pd \O} \pd_{\o} T_\o|_{\o = \o_0} [u_0] {u}_0 \, d\sigma \\
\nm
&& \displaystyle + \frac{1}{\e_m} \int_{\pd B_\delta} ({u}_0 \dfrac{\pd u_1}{\pd \nu} - u_1 \dfrac{\pd {u}_0}{\pd \nu}) \, d\sigma. \end{array}$$
From (\ref{et2}), we have
$$\int_{\pd B_\delta} ({u}_0 \dfrac{\pd u_1}{\pd \nu} - u_1 \dfrac{\pd {u}_0}{\pd \nu} )  \, d\sigma \xrarr{\delta \to 0} - M(\varepsilon_m/ \varepsilon_c,B) \nabla u_0(z) \cdot \nabla {u}_0(z) .$$
Therefore,
$$-\l_1 \mu_m \int_\Omega u_0^2 \, dx - \dfrac{\l_1}{2 \o_0} \int_{\pd \O} \pd_\o  T_\o |_{\o = \o_0}[u_0] {u}_0\, d\sigma  =  - \frac{1}{\e_m} M(\varepsilon_m/ \varepsilon_c, B) \nabla u_0(z) \cdot \nabla {u}_0 (z),  $$
and finally, we arrive at
\begin{equation}
\label{finaleq} \displaystyle \l_1 = \dfrac{ M(\varepsilon_m/ \varepsilon_c, B) \nabla u_0(z) \cdot \nabla {u}_0 (z) }{\e_m \mu_m \int_\Omega u_0^2 \, dx + (1/(2\o_0)) \,  \e_m \int_{\pd \O} \pd_\o T_\o  |_{\o = \o_0} [u_0] {u}_0\, d\sigma } ,\end{equation}
or equivalently,
\begin{equation}
\label{finaleqeq} \displaystyle \omega_1 = \dfrac{ M(\varepsilon_m/ \varepsilon_c, B) \nabla u_0(z) \cdot \nabla {u}_0 (z) }{2\o_0 \mu_m \e_m \int_\Omega u_0^2 \, dx +  \,  \e_m \int_{\pd \O} \pd_\o T_\o  |_{\o = \o_0} [u_0] {u}_0\, d\sigma } . \end{equation}
In the multi-dimensional case, the effect of radiation on the shift of the scattering resonance $\omega_0$ is given by
$\e_m  \int_{\pd \O} \pd_\o T_\o  |_{\o = \o_0} [u_0] {u}_0\, d\sigma$.
Note also that formula (\ref{finaleqeq}) reduces to (\ref{eq:onedimensionalpartial}) in the one-dimensional case. In fact, the polarization tensor $M$ reduces to $\alpha$ defined by (\ref{defalpha}) and the operator $T_\omega$ corresponds to the multiplication by $- i \omega$ at $a$ and $+ i \omega$ at $b$. 
If one relaxes the assumption $\mu_c=\mu_m$, one can easily generalize formula \eqref{finaleqeq} by computing, as in \cite{khelifi} and in Section \ref{sec-1}, the second-order corrector $e_2$. We then get the desired result. \cqfd
% On the other hand, formula (\ref{finaleqeq}) can be easily generalized to the following scattering resonance perturbation problem:
%$$\begin{cases}
%(1/\varepsilon_m) \Delta u_\delta + \o_\delta^2 \e_m u_\delta = 0 & \mbox{in }  \O \smin \bar D, \\
%(1/\varepsilon_c) \Delta u_\delta + \o_\delta^2 \e_c u_\delta = 0 & \mbox{in }  D, \\
%(1/\varepsilon_m) \dfrac{\pd u_\delta}{\pd \nu} \big|_{+} = (1/\varepsilon_c) \dfrac{\pd u_\delta}{\pd \nu} \big|_{-}  & \mbox{on }  \pd D, \\
%(1/\varepsilon_m) \dfrac{\pd u_\delta}{\pd \nu} = T_{\o_\delta} [u_\delta] & \mbox{on }  \pd \O.
%\end{cases}$$
%By computing, as in \cite{khelifi}, the second-order corrector $e_2$, we can show that the result holds. \cqfd

\section{Perturbations of whispering-gallery modes by an external particle} \label{sec-3}

\begin{figure}[h!]
\def\svgwidth{0.9\linewidth}
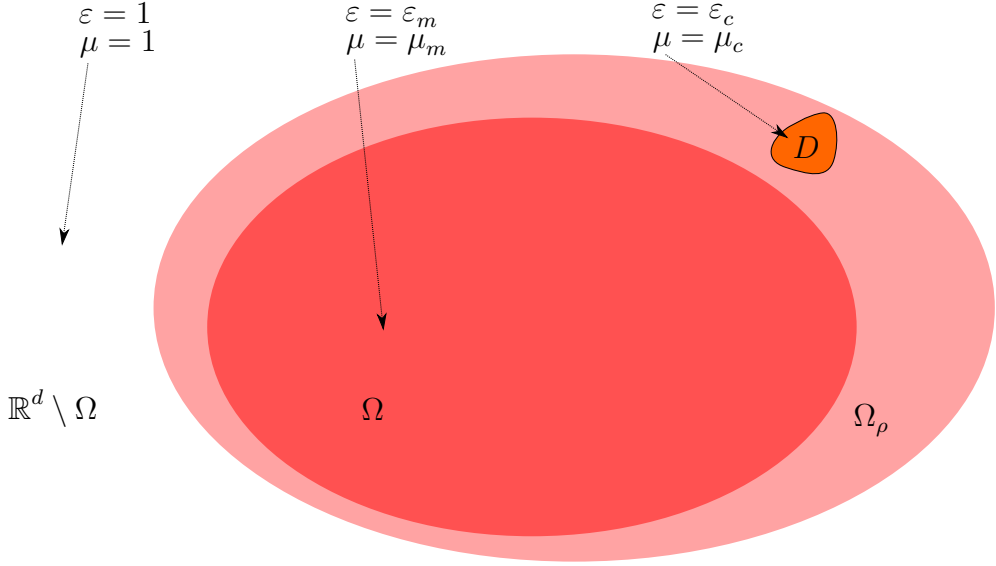
\caption{\footnotesize{Perturbed cavity by an external particle.}}
\label{figadd3}
\end{figure}
Whispering-gallery modes are modes which are confined near the boundary of the cavity. Their existence can be proved analytically or by a boundary layer approach based on WKB (high frequency) asymptotics \cite{agallery1, lam, ludwig, matkowsky, WG1, GW2,GW3}.  Whispering-gallery modes are exploited to probe the local surroundings \cite{haroche2,haroche3,agallery2}. Biosensors based on the shift of whispering-gallery modes in open cavities by small particles have been also  described by use of Bethe-Schwinger type formulas, where the effect of radiation is neglected \cite{WG0,WG1,WG4,WG5}. In this section, we provide a generalization of the formula derived in the previous section and discuss its validity for whispering-gallery modes.

Assume that $\omega_0$ is a whispering-gallery mode of the open cavity $\Omega$. Let $\Omega_\rho$ be a small neighborhood of $\Omega$. Suppose that the particle $D$ is in $\Omega_\rho \setminus \overline \Omega$, see Figure \ref{figadd3}. If the characteristic size $\delta$ of $D$ is much smaller than $\rho$, which is in turn much smaller than $2\pi/(\sqrt{\e_m \mu_m}\omega_0)$, then by the same arguments as those in the previous section, the leading-order term in the shift of the resonant frequency $\omega_0$ is given by
$$\omega_1 \simeq \dfrac{  M(1/\varepsilon_c, B) \nabla v_0(z) \cdot \nabla {v}_0 (z)  +  \o_0^2 |B| (\mu_c -1) (v_0(z))^2}{2\omega_0 \mu_m \e_m \int_\Omega u_0^2 \, dx +  \e_m \int_{\pd \O} \pd_\o T_\o  |_{\o = \o_0} [u_0] {u}_0\, d\sigma } .$$
Here, the polarization tensor $M(\varepsilon_m/\varepsilon_c, B)$ in (\ref{finaleq}) is replaced by $M(1/\varepsilon_c, B)$ since $\varepsilon$ in the medium surrounding the particle is equal to $1$ and $v_0$ is defined in $\mathbb{R}^d$ by 
\begin{equation} \label{defv0p1}
v_0(x) = -\omega_0^2  (\mu_m -1) \int_{\Omega} \Gamma(x-y;\omega_0) u_0(y)\, dy + (\frac{1}{\e_m} -1) \int_\Omega  \nabla_y \Gamma(x-y;\omega_0) \cdot \nabla u_0(y)\, dy, 
\end{equation}
where $\Gamma(\cdot;\omega_0)$ is the fundamental solution of $\Delta +\omega_0^2$, which satisfies the outgoing radiation condition. 
 We remark that $v_0 =u_0$ in $\Omega$. Moreover,  the assumption that $\omega_0$ is a whispering-gallery mode is needed in order to have the gradient of $v_0$ at the location of the particle to have a significant magnitude.

Now, assume that the particle $D$ is plasmonic, i.e., $\varepsilon_c$ depends on the frequency $\omega$ and can take negative values. In this case, there is a discrete set of frequencies, called plasmonic resonant frequencies, such that at these frequencies problem (\ref{pbv}) is nearly singular, and therefore the  polarization tensor associated with the particle $D$ blows up at those frequencies, see \cite{plasmonic1,plasmonic2,plasmonic3}. Assume that the plasmonic particle is coupled to the cavity, i.e., there is a whispering-gallery cavity mode $\omega_0$ such that  $\Re \omega_0$ is a plasmonic resonance of the particle.

 Then when the particle $D$ is illuminated at the frequency $\Re \omega_0$, its effect on the cavity mode $\omega_0$ is given by the following proposition. 
 \begin{proposition} \label{propGW} We have
\begin{equation}
\label{finaleq3} \displaystyle \omega_1 \simeq \dfrac{ M((1/\varepsilon_c) (\Re \omega_0), B) \nabla v_0(z) \cdot \nabla {v}_0 (z) +  \o_0^2 |B|  (\mu_c -1) (v_0(z))^2}{2\omega_0 \mu_m \e_m \int_\Omega u_0^2 \, dx +  \, \e_m  \int_{\pd \O} \pd_\o T_\o  |_{\o = \o_0} [u_0] {u}_0\, d\sigma } ,\end{equation} where $v_0$ is defined by (\ref{defv0p1}). 
\end{proposition}

Proposition \ref{propGW} shows that despite their small size, plasmonic particles significantly change the cavity modes when their plasmonic resonances are close to the cavity modes.

Finally, suppose that $\omega_0$ is of multiplicity $m$. Then, following \cite{triki,myself1,myself4},  $\omega_0$ can be split into $m$  scattering resonances $\o_{\delta,j}$ having the following approximations:
\begin{align}
\label{eq:open:asympformula:mult}
\o_{\delta,j}^2 \simeq \o_0^2  + {\delta^d} \eta_j,
\end{align}
with $\eta_j$ being the $j$-th eigenvalue of the matrix
\begin{equation} \label{eq:close:cond:eigasympformula:varmatrix}
\left( \frac{  M \nabla v_{0,p}(z) \cdot \nabla {v}_{0,q} (z)  +  \o_0^2 |B|  (\mu_c -1) v_{0,p}(z) v_{0,q}(z)}{\mu_m \e_m \int_\Omega u_{0,p} u_{0,q} \, dx + (1/(2\o_0)) \, \e_m \int_{\pd \O} \pd_\o T_\o  |_{\o = \o_0} [u_{0,q}] {u}_{0,p}\, d\sigma } \right)_{p,q=1}^m.
\end{equation}
Here, $\{ v_{0,q}\}_{q=1,\ldots,m}$ are obtained by (\ref{defv0p1} with $\{ u_{0,q}\}_{q=1,\ldots,m}$ being an orthonormal eigenspace associated with $\omega_0$.

\section{Numerical illustrations} \label{sec-4}
In two dimensions, when the cavity and the small-volume particle are disks we can use the multipole expansion method to efficiently compute the perturbations of the whispering-gallery modes \cite{martin2006multiple}. Our approach is as follows. We first use a projective eigensolver \cite{berljafa2014rational} to obtain a coarse estimate of the locations of the  resonances of a two disk system. We then focus on the particular resonances in this set that correspond to the whispering-gallery modes of the open cavity and obtain a refined estimate of their locations using Muller's method \cite{ammari2017mathematical}.

It is well-known that boundary integral formulations of the exterior and transmission scattering problems are prone to so-called spurious resonances which can interfere with the search for the true scattering resonances \cite{brakhage1965dirichletsche}. In order to achieve a better separation between the spurious resonances and the true resonances when using the projective eigensolver,  a combined field integral equation approach can be used \cite{rapun2006indirect, steinbach2017combined}.

\begin{figure}[!htbp]
\begin{center}
\includegraphics[scale=.6]{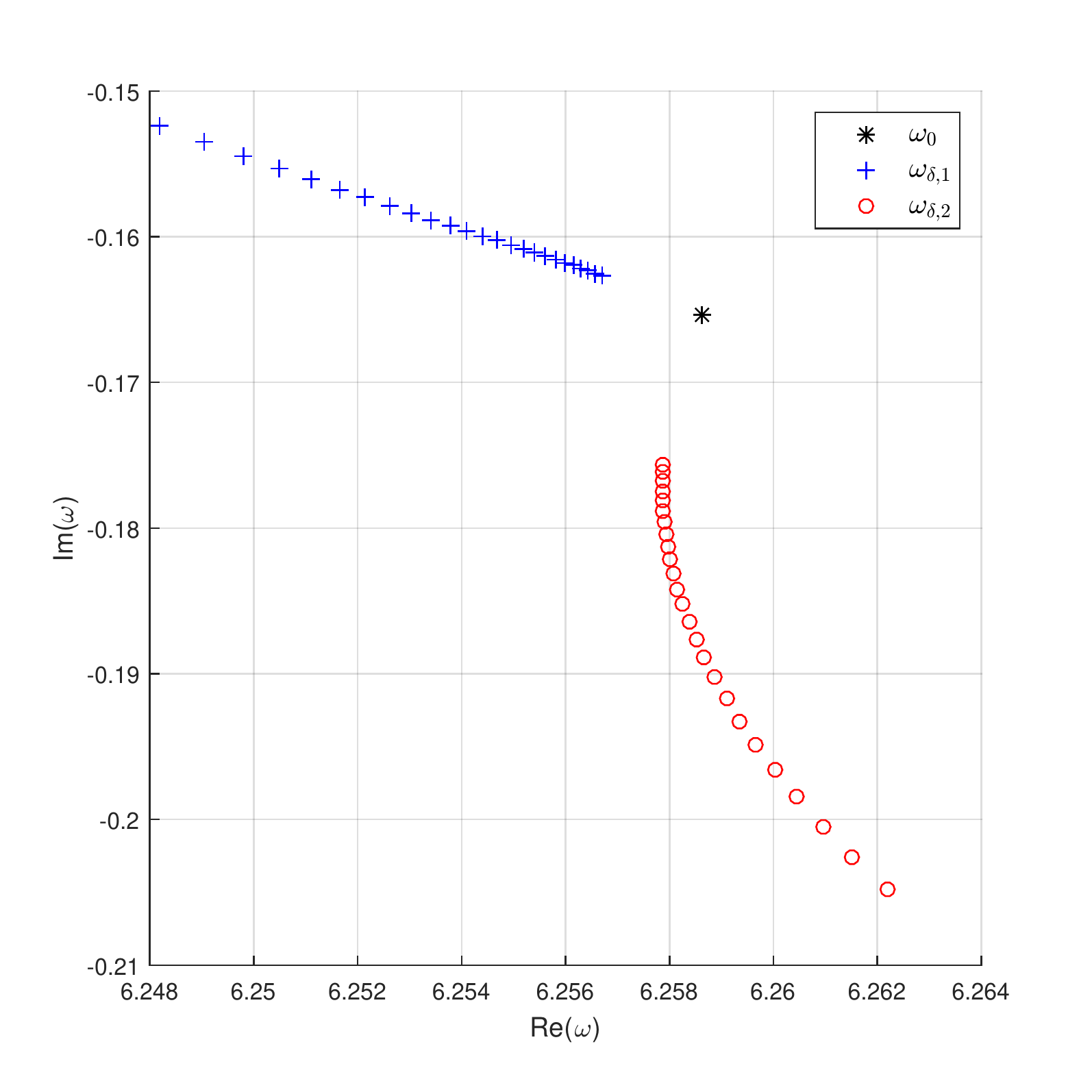} \hspace{10pt}
\end{center}
\caption{\footnotesize{As the size of the small disk $\delta \to 0$, the perturbed whispering-gallery modes $\omega_{\delta,1}$ and $\omega_{\delta,2}$ converge towards the unperturbed mode $\omega_0$.}}
\label{fig:open:sizechange}
\end{figure}

\begin{figure}[!htbp]
\begin{center}
\includegraphics[scale=.5]{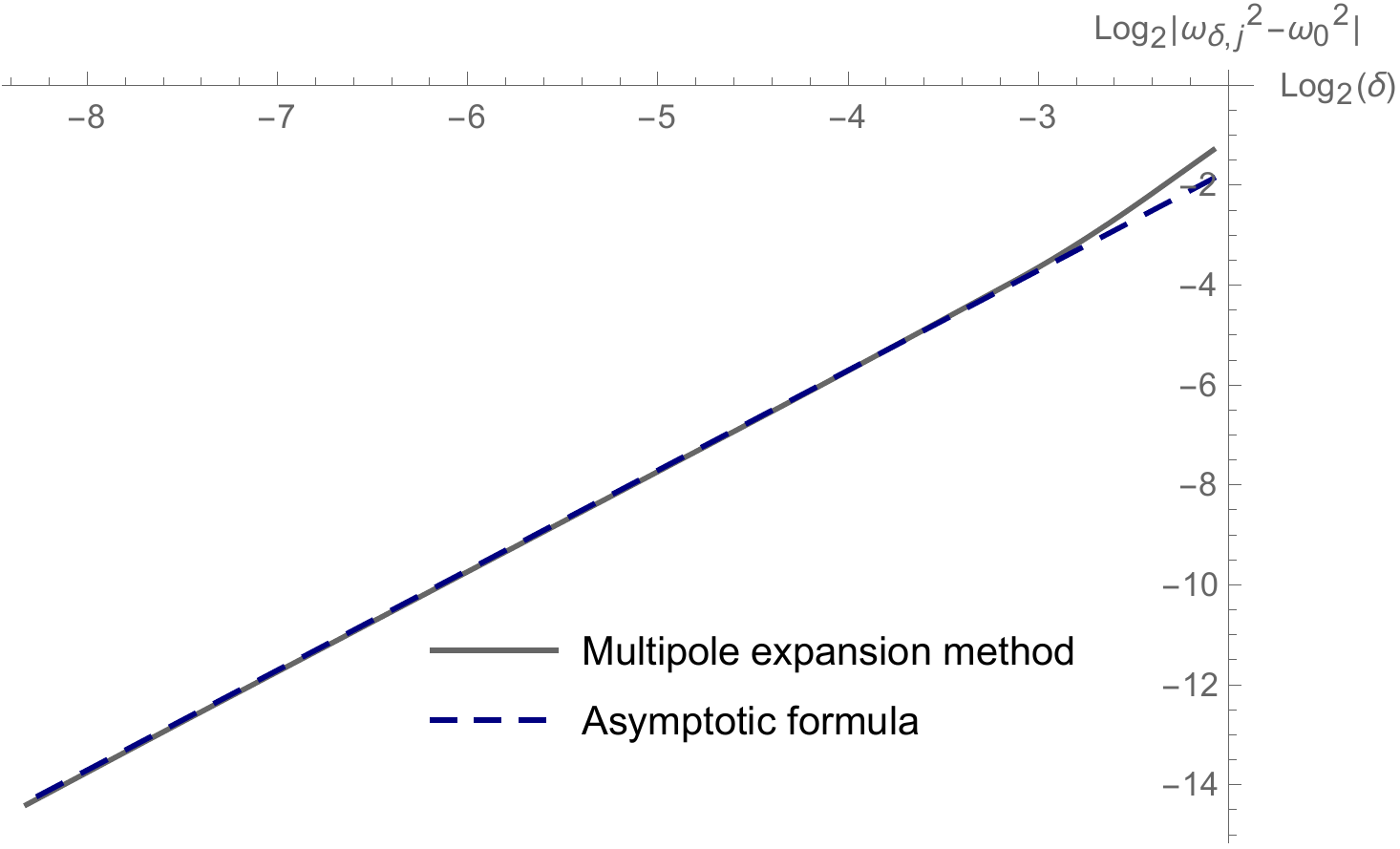}
\end{center}
\caption{\footnotesize{Comparison between the asymptotic formula for the perturbation $|\omega_{\delta,1}^2 - \omega_0^2|$ of the whispering-gallery mode and the perturbation computed numerically as the size of the small disk $\delta \to 0$.}}
\label{fig:open:sizechange2}
\end{figure}

Throughout this section, $\O$ is a disk of radius $1$ centered at the origin and $\omega_0$ is the frequency of a whispering-gallery mode. Let  $D$ be a disk of radius $\delta$ centered at $(1+2\delta, 0)$. Suppose  that $\varepsilon_m = \varepsilon_c = 1/5$.
The behavior of $\o_{\delta,1},\o_{\delta,2}$ as $\delta \to 0$ is plotted in Figure \ref{fig:open:sizechange}. Formula \eqref{eq:open:asympformula:mult} matches the behavior of the eigenvalue perturbation as can be seen in Figure \ref{fig:open:sizechange2}. On the other hand, we can easily reconstruct $\delta$ from a single scattering resonance shift.

\begin{figure}[!htbp]
\begin{center}
\includegraphics[scale=.5]{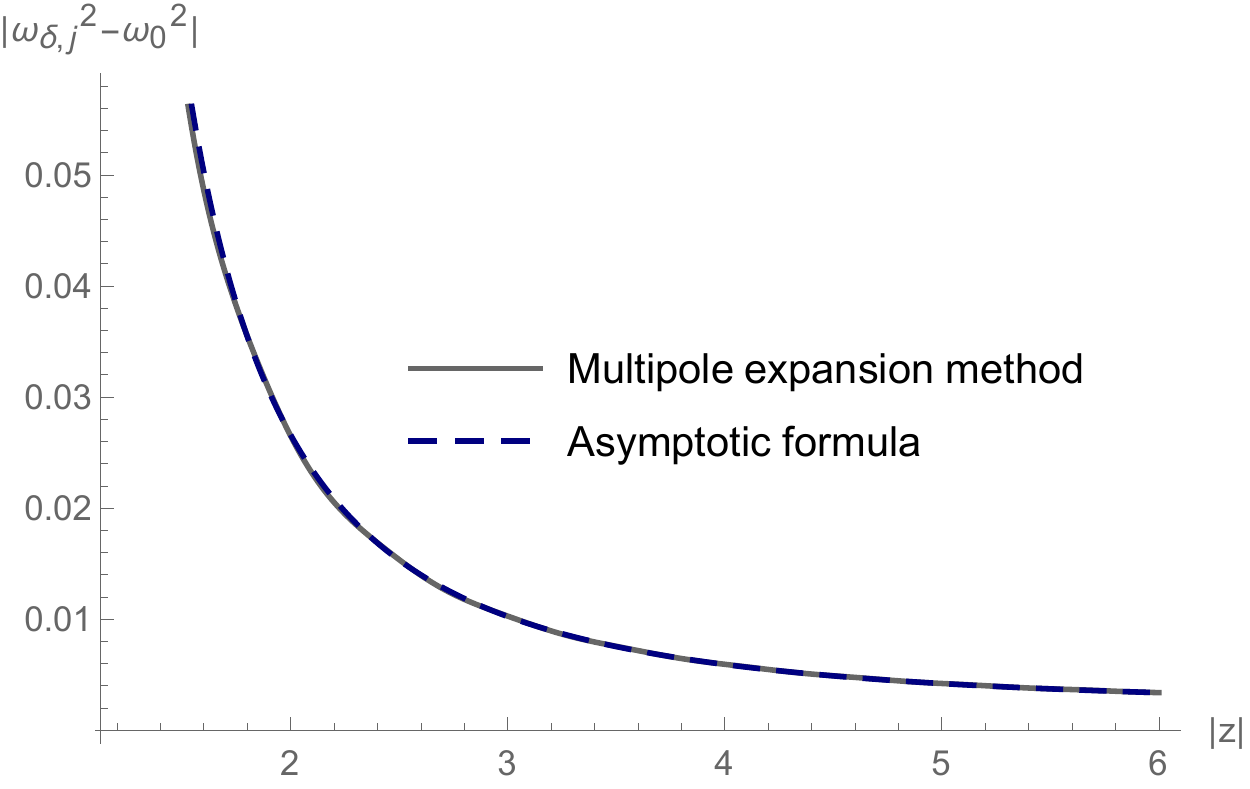} \hspace{10pt}
\includegraphics[scale=.5]{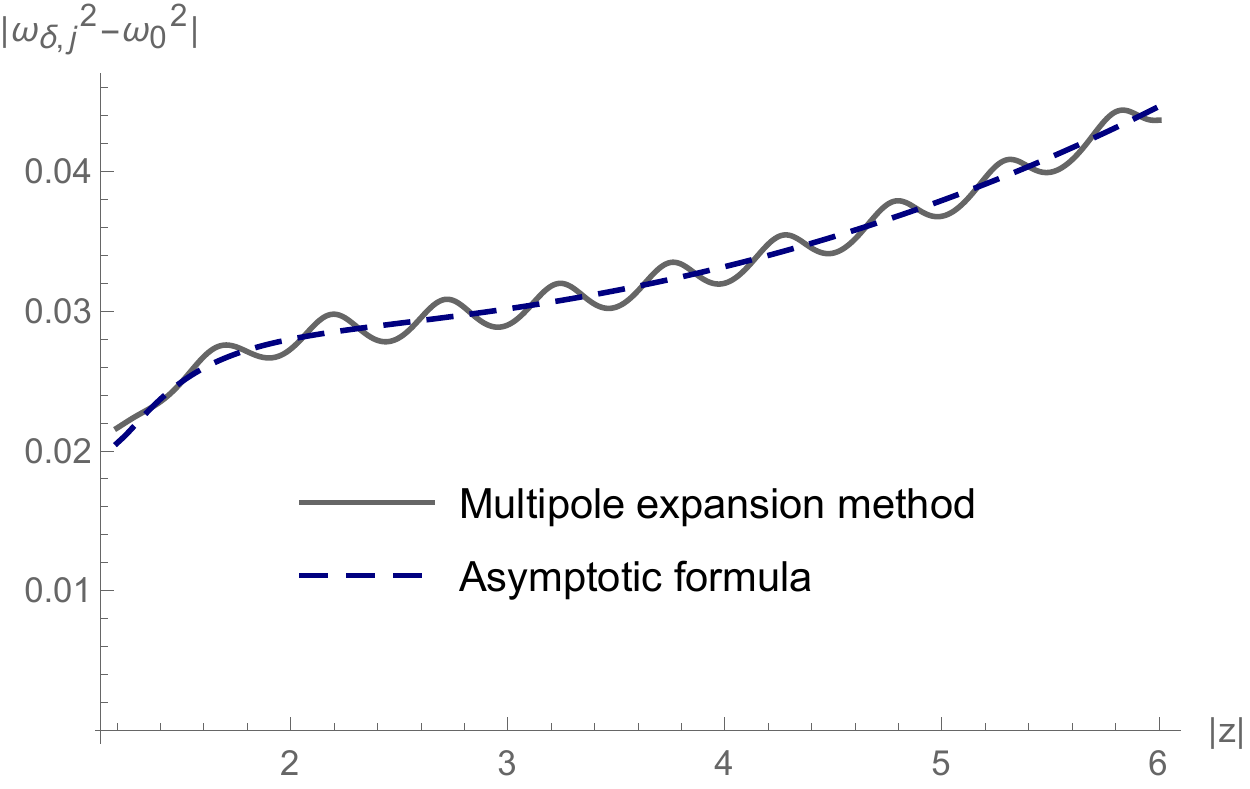}
\end{center}
\caption{\footnotesize{Comparison between the asymptotic formula for the perturbation $|\omega_{\delta,j}^2 - \omega_0^2|$ of the whispering-gallery mode and the perturbation computed numerically as the position of the inclusion $(z,0)$ varies. The plot on the left corresponds to the perturbed resonance $\omega_{\delta,1}$ and the plot on the right corresponds to the perturbed resonance $\omega_{\delta,2}$.}}
\label{fig:open:positionchange}
\end{figure}

Next, consider a disk $D_\delta$ of radius $\delta = 0.1$ centered at $(z, 0)$. A plot of $|\o_{\delta,j}^2 - \o_0^2|$ as $z$ varies between $1.2$ and $6$ is presented in Figure \ref{fig:open:positionchange}.

%Now, consider a disk $D_\e$ of radius $\e$ centered at $(1+ \e, 0)$.
%A plot of $|\o_{\e,j}^2 - \o_0^2|$ as $\e \to 0$, i.e., as
%the particle radius shrinks while concurrently becoming closer to $\O$, with both changes occurring at the same speed, is presented in \Cref{fig:open:position+sizechange}.
%\begin{figure}[!htbp]
%\begin{center}
%\includegraphics[scale=.5]{fig-p1/Open_PositionAndSizeChange_1.pdf} \hspace{10pt}
%\includegraphics[scale=.5]{fig-p1/Open_PositionAndSizeChange_2.pdf}
%\end{center}
%\caption{Resonance perturbation $|\omega_{\varepsilon,j}^2 - \omega_0^2|$ as the size of the small particle $D$ shrinks while concurrently becoming closer to $\Omega$, with both changes occurring at the same speed. The plot on the left corresponds to the perturbed resonance $\omega_{\varepsilon,1}$ and the plot on the right corresponds to the perturbed resonance $\omega_{\varepsilon,2}$.}
%\label{fig:open:position+sizechange}
%\end{figure}

By using (\ref{eq:open:asympformula:mult}), one can also reconstruct the polarization tensor. We highlight here the case of plasmonic particles. In this case we have a strong enhancement in the frequency shift, which allows for the recognition of much smaller particles.

\begin{figure}[!htbp]
\begin{center}
\includegraphics[scale=.6]{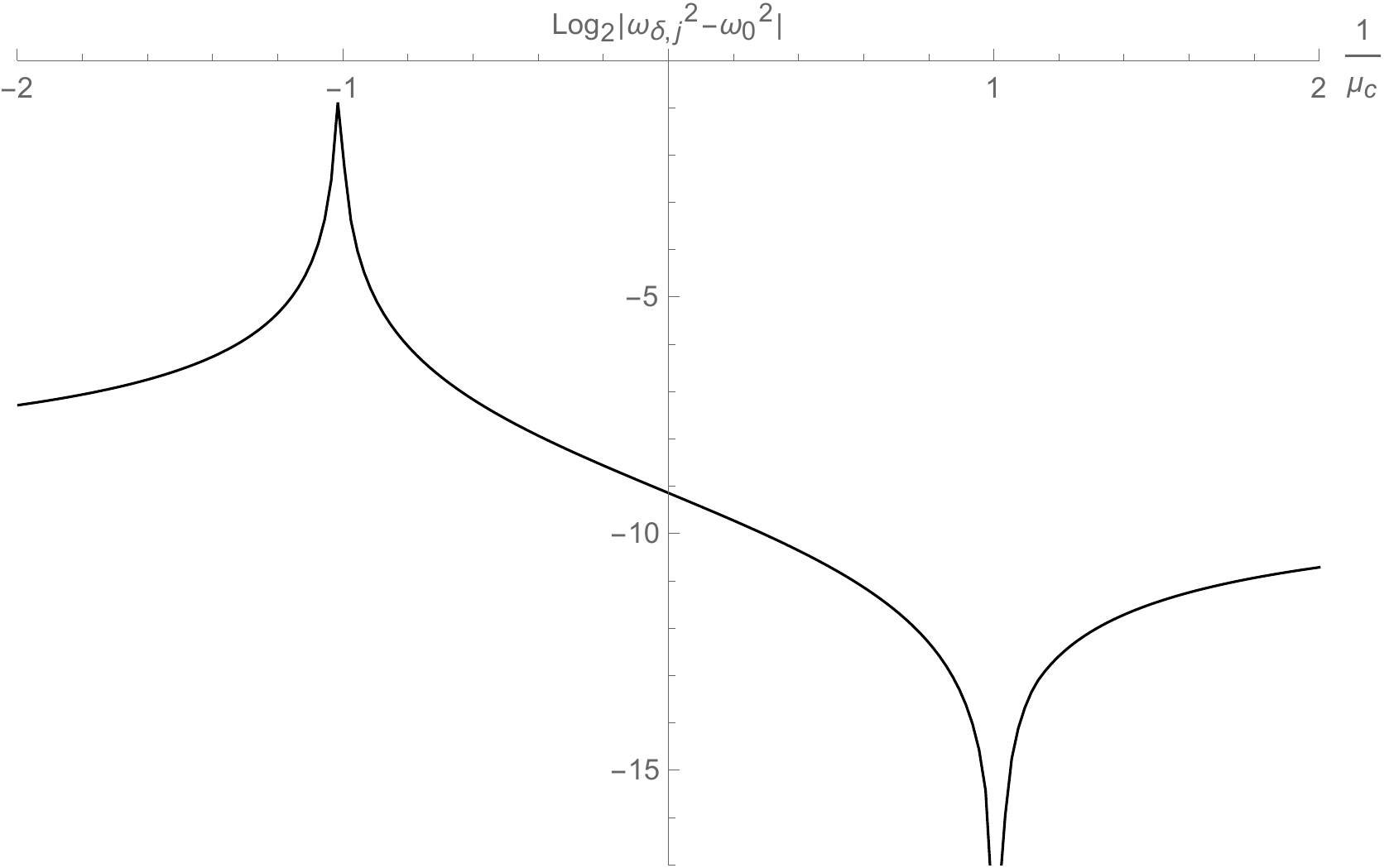}
\end{center}
\caption{\footnotesize{Resonance perturbation $|\omega_{\delta,1}^2 - \omega_0^2|$ as a function of $1/\varepsilon_c$, here allowed to also take negative values.}}
\label{fig:Open_CondChange}
\end{figure}

Consider a  disk $D$ of radius $0.1$ centered at $(1.2, 0)$. Suppose $\varepsilon_m =1/5$.
A plot of $|\o_{\delta,1}^2 - {\o_0}^2|$ as $1/\varepsilon_c$ varies is presented in Figure \ref{fig:Open_CondChange}.
Notice the high peak in the perturbation as $\varepsilon_c$ approaches the value $-1$.

Finally, suppose we have $n$ particles arranged outside $\O$ as vertices of a regular $n$-gon, and tangent to $\pd \O$.
Suppose all the particles have the same polarization tensor $M$.
As $\delta \to 0$, we can consider the contribution of each particle independently, and thus summing up \eqref{eq:open:asympformula:mult} we have
\begin{align}
\label{eq:open:counting}
\o_{\delta,j}^2 - \o_0^2 \simeq  \sum_{i=1}^n \delta^d \eta_{i,j},
\end{align}
where $\eta_{i,j}$ is the $j$-th eigenvalue of \eqref{eq:close:cond:eigasympformula:varmatrix} with $z$ substituted by $z_i $, the center of the $i$-th particle.
Considering different frequencies, we can reconstruct $n$ by looking for a minimizer of an appropriate discrepancy functional.

\section{Concluding remarks} \label{sec-5}
In this paper, the leading-order term in the shifts of scattering resonances by small particles is derived and the effect of radiation on the perturbations of open cavity modes is characterized. The formula is in terms of the position and the polarization tensor of the particle. It is valid for arbitrary-shaped particles. By reconstructing the polarization tensor of the small particle, the orientation of the perturbing particle can be inferred, which affords the possibility of orientational binding studies in biosensing. It is also worth mentioning that, based on \cite{volkov, lee}, the derived formula can be generalized to open electromagnetic and elastic cavities.

\end{document}

%% file: 1Dbounded.pdf_tex
%% Creator: Inkscape 0.91_64bit, www.inkscape.org
%% Accompanies image file '1Dbounded.pdf' (pdf, eps, ps)
\begingroup%
  \makeatletter%
  \providecommand\color[2][]{%
    \errmessage{(Inkscape) Color is used for the text in Inkscape, but the package 'color.sty' is not loaded}%
    \renewcommand\color[2][]{}%
  }%
  \providecommand\transparent[1]{%
    \errmessage{(Inkscape) Transparency is used (non-zero) for the text in Inkscape, but the package 'transparent.sty' is not loaded}%
    \renewcommand\transparent[1]{}%
  }%
  \providecommand\rotatebox[2]{#2}%
  \ifx\svgwidth\undefined%
    \setlength{\unitlength}{841.88976378bp}%
    \ifx\svgscale\undefined%
      \relax%
    \else%
      \setlength{\unitlength}{\unitlength * \real{\svgscale}}%
    \fi%
  \else%
    \setlength{\unitlength}{\svgwidth}%
  \fi%
  \global\let\svgwidth\undefined%
  \global\let\svgscale\undefined%
  \makeatother%
  \begin{picture}(1,0.70707071)%
    \put(0,0){\includegraphics[width=\unitlength,page=1]{1Dbounded.pdf}}%
    \put(0.28962638,0.37999551){\color[rgb]{0,0,0}\makebox(0,0)[lb]{\smash{$a$}}}%
    \put(0.57030751,0.37999551){\color[rgb]{0,0,0}\makebox(0,0)[lb]{\smash{$b$}}}%
    \put(0.39496353,0.50242343){\color[rgb]{0,0,0}\makebox(0,0)[lb]{\smash{$\varepsilon_m$}}}%
    \put(0.83495497,0.38618616){\color[rgb]{0,0,0}\makebox(0,0)[lb]{\smash{$x$}}}%
    \put(0.24243321,0.59423081){\color[rgb]{0,0,0}\makebox(0,0)[lb]{\smash{The unperturbed cavity}}}%
    \put(0,0){\includegraphics[width=\unitlength,page=2]{1Dbounded.pdf}}%
    \put(0.65176217,0.34730646){\color[rgb]{0,0,0}\makebox(0,0)[lb]{\smash{Impedance boundary conditions}}}%
    \put(0,0){\includegraphics[width=\unitlength,page=3]{1Dbounded.pdf}}%
    \put(0.28539091,0.05493591){\color[rgb]{0,0,0}\makebox(0,0)[lb]{\smash{$a$}}}%
    \put(0.56417158,0.05493591){\color[rgb]{0,0,0}\makebox(0,0)[lb]{\smash{$b$}}}%
    \put(0.30253858,0.19104955){\color[rgb]{0,0,0}\makebox(0,0)[lb]{\smash{$\varepsilon_m$}}}%
    \put(0.8307195,0.06112656){\color[rgb]{0,0,0}\makebox(0,0)[lb]{\smash{$x$}}}%
    \put(0,0){\includegraphics[width=\unitlength,page=4]{1Dbounded.pdf}}%
    \put(0.64907436,0.00986586){\color[rgb]{0,0,0}\makebox(0,0)[lb]{\smash{Impedance boundary conditions}}}%
    \put(0,0){\includegraphics[width=\unitlength,page=5]{1Dbounded.pdf}}%
    \put(0.47484267,0.07245469){\color[rgb]{0,0,0}\makebox(0,0)[lb]{\smash{$\frac{\delta}{2}$}}}%
    \put(0.38120999,0.07168091){\color[rgb]{0,0,0}\makebox(0,0)[lb]{\smash{$-\frac{\delta}{2}$}}}%
    \put(0.44179661,0.19107801){\color[rgb]{0,0,0}\makebox(0,0)[lb]{\smash{$\varepsilon_c$}}}%
    \put(0.39697116,0.47801093){\color[rgb]{0,0,0}\makebox(0,0)[lb]{\smash{$\mu_m$}}}%
    \put(0.3029707,0.16893031){\color[rgb]{0,0,0}\makebox(0,0)[lb]{\smash{$\mu_m$}}}%
    \put(0.44183873,0.1691893){\color[rgb]{0,0,0}\makebox(0,0)[lb]{\smash{$\mu_c$}}}%
    \put(0,0){\includegraphics[width=\unitlength,page=6]{1Dbounded.pdf}}%
    \put(0.24223592,0.24742486){\color[rgb]{0,0,0}\makebox(0,0)[lb]{\smash{The perturbed cavity}}}%
  \end{picture}%
\endgroup%

%% file: 2Dinternal.pdf_tex
%% Creator: Inkscape 0.91_64bit, www.inkscape.org
%% Accompanies image file '2Dinternal.pdf' (pdf, eps, ps)
\begingroup%
  \makeatletter%
  \providecommand\color[2][]{%
    \errmessage{(Inkscape) Color is used for the text in Inkscape, but the package 'color.sty' is not loaded}%
    \renewcommand\color[2][]{}%
  }%
  \providecommand\transparent[1]{%
    \errmessage{(Inkscape) Transparency is used (non-zero) for the text in Inkscape, but the package 'transparent.sty' is not loaded}%
    \renewcommand\transparent[1]{}%
  }%
  \providecommand\rotatebox[2]{#2}%
  \ifx\svgwidth\undefined%
    \setlength{\unitlength}{841.88976378bp}%
    \ifx\svgscale\undefined%
      \relax%
    \else%
      \setlength{\unitlength}{\unitlength * \real{\svgscale}}%
    \fi%
  \else%
    \setlength{\unitlength}{\svgwidth}%
  \fi%
  \global\let\svgwidth\undefined%
  \global\let\svgscale\undefined%
  \makeatother%
  \begin{picture}(1,0.70707071)%
    \put(0,0){\includegraphics[width=\unitlength,page=1]{2Dinternal.pdf}}%
    \put(0.36214912,0.15553844){\color[rgb]{0,0,0}\makebox(0,0)[lb]{\smash{$\Omega$}}}%
    \put(0.69257151,0.26232597){\color[rgb]{0,0,0}\makebox(0,0)[lb]{\smash{$D$}}}%
    \put(0.34744645,0.51691373){\color[rgb]{0,0,0}\makebox(0,0)[lb]{\smash{$\varepsilon=\varepsilon_m$}}}%
    \put(0.34822026,0.49137755){\color[rgb]{0,0,0}\makebox(0,0)[lb]{\smash{$\mu=\mu_m$}}}%
    \put(0.62896307,0.51933228){\color[rgb]{0,0,0}\makebox(0,0)[lb]{\smash{$\varepsilon=\varepsilon_c$}}}%
    \put(0.62973685,0.49379612){\color[rgb]{0,0,0}\makebox(0,0)[lb]{\smash{$\mu=\mu_c$}}}%
    \put(0.10276349,0.51455306){\color[rgb]{0,0,0}\makebox(0,0)[lb]{\smash{$\varepsilon=1$}}}%
    \put(0.1035373,0.48901691){\color[rgb]{0,0,0}\makebox(0,0)[lb]{\smash{$\mu=1$}}}%
    \put(0,0){\includegraphics[width=\unitlength,page=2]{2Dinternal.pdf}}%
    \put(0.03791733,0.15631227){\color[rgb]{0,0,0}\makebox(0,0)[lb]{\smash{$\mathbb{R}^d\setminus\Omega$}}}%
    \put(0.16559809,0.60822482){\color[rgb]{0,0,0}\makebox(0,0)[lb]{\smash{Cavity perturbed by an internal particle}}}%
  \end{picture}%
\endgroup%

%% file: 2Dexternal.pdf_tex
%% Creator: Inkscape 0.91_64bit, www.inkscape.org
%% Accompanies image file '2Dexternal.pdf' (pdf, eps, ps)
\begingroup%
  \makeatletter%
  \providecommand\color[2][]{%
    \errmessage{(Inkscape) Color is used for the text in Inkscape, but the package 'color.sty' is not loaded}%
    \renewcommand\color[2][]{}%
  }%
  \providecommand\transparent[1]{%
    \errmessage{(Inkscape) Transparency is used (non-zero) for the text in Inkscape, but the package 'transparent.sty' is not loaded}%
    \renewcommand\transparent[1]{}%
  }%
  \providecommand\rotatebox[2]{#2}%
  \ifx\svgwidth\undefined%
    \setlength{\unitlength}{841.88976378bp}%
    \ifx\svgscale\undefined%
      \relax%
    \else%
      \setlength{\unitlength}{\unitlength * \real{\svgscale}}%
    \fi%
  \else%
    \setlength{\unitlength}{\svgwidth}%
  \fi%
  \global\let\svgwidth\undefined%
  \global\let\svgscale\undefined%
  \makeatother%
  \begin{picture}(1,0.70707071)%
    \put(0,0){\includegraphics[width=\unitlength,page=1]{2Dexternal.pdf}}%
    \put(0.36214912,0.15553844){\color[rgb]{0,0,0}\makebox(0,0)[lb]{\smash{$\Omega$}}}%
    \put(0.75823255,0.39802546){\color[rgb]{0,0,0}\makebox(0,0)[lb]{\smash{$D$}}}%
    \put(0.34744645,0.52451569){\color[rgb]{0,0,0}\makebox(0,0)[lb]{\smash{$\varepsilon=\varepsilon_m$}}}%
    \put(0.34822026,0.49897951){\color[rgb]{0,0,0}\makebox(0,0)[lb]{\smash{$\mu=\mu_m$}}}%
    \put(0.62896307,0.52693424){\color[rgb]{0,0,0}\makebox(0,0)[lb]{\smash{$\varepsilon=\varepsilon_c$}}}%
    \put(0.62973685,0.50139808){\color[rgb]{0,0,0}\makebox(0,0)[lb]{\smash{$\mu=\mu_c$}}}%
    \put(0.10276349,0.52215502){\color[rgb]{0,0,0}\makebox(0,0)[lb]{\smash{$\varepsilon=1$}}}%
    \put(0.1035373,0.49661887){\color[rgb]{0,0,0}\makebox(0,0)[lb]{\smash{$\mu=1$}}}%
    \put(0,0){\includegraphics[width=\unitlength,page=2]{2Dexternal.pdf}}%
    \put(0.03791733,0.15631227){\color[rgb]{0,0,0}\makebox(0,0)[lb]{\smash{$\mathbb{R}^d\setminus\Omega$}}}%
    \put(0.16559809,0.60822482){\color[rgb]{0,0,0}\makebox(0,0)[lb]{\smash{Cavity perturbed by an external particle}}}%
    \put(0.81419706,0.14992608){\color[rgb]{0,0,0}\makebox(0,0)[lb]{\smash{$\Omega_\rho$}}}%
  \end{picture}%
\endgroup%

%% file: paper_1_final_TM.bbl
\begin{thebibliography}{2}

\bibitem{paper2} H. Ammari, A. Dabrowski, B. Fitzpatrick, and P. Millien,
Perturbations of the scattering resonances of an open cavity by small particles. Part II: The transverse electric polarization case, submitted.

\bibitem{plasmonic1} H. Ammari, Y. Deng, and P. Millien, Surface plasmon resonance of nanoparticles and applications in imaging, Arch. Ration. Mech. Anal., 220 (2016),
109--153.

\bibitem{ammari2017mathematical} H. Ammari, B. Fitzpatrick, H. Kang, M. Ruiz, S. Yu, and H. Zhang, \textsl{Mathematical and computational methods in photonics and phononics}, Mathematical Surveys and Monographs, Vol. 235, American Mathematical Society, Providence, 2018.

\bibitem{kang_book} H. Ammari and H. Kang,  \textsl{Polarization and moment tensors. With applications to inverse problems and effective medium theory},  Applied Mathematical Sciences, 162. Springer, New York, 2007.

\bibitem{lee}  H. Ammari, H. Kang, and H. Lee, Asymptotic expansions for eigenvalues of the Lamé system in the presence of small inclusions, Comm. Partial Differential Equations, 32 (2007), 1715--1736.

\bibitem{khelifi} H. Ammari and A. Khelifi, Electromagnetic scattering by small dielectric
inhomogeneities, J. Math. Pures Appl., 82 (2003), 749--842.

\bibitem{AmmariMillien2018} H. Ammari and P. Millien,
Shape and size dependence of dipolar plasmonic resonance of nanoparticles, J. Math. Pures Appl., to appear (arXiv:1804.11092).

\bibitem{plasmonic2} H.  Ammari, P. Millien, M. Ruiz, and H. Zhang, Mathematical analysis of plasmonic nanoparticles: the scalar case, Arch. Ration. Mech. Anal., 224 (2017),  597--658.


\bibitem{moskow} H. Ammari and S. Moskow, Asymptotic expansions for eigenvalues in the presence of small inhomogeneities, Math. Methods Appl. Sci., 26 (2003),  67--75.

\bibitem{nedelec} H. Ammari and J.-C. N\'ed\'elec, Full low-frequency asymptotics for the reduced wave equation, Appl. Math. Lett., 12 (1999), 127--131.

\bibitem{plasmonic3} H. Ammari, M. Ruiz, S. Yu, and H. Zhang, Mathematical analysis of plasmonic resonances for nanoparticles: the full Maxwell equations, J. Differential Equations 261 (2016), 3615--3669.


\bibitem{triki}  H. Ammari and F. Triki, Splitting of resonant and scattering frequencies under shape deformation, J. Differential Equations,  202 (2004), 231--255.

\bibitem{volkov} H. Ammari and D. Volkov, Asymptotic formulas for perturbations in the eigenfrequencies of the full Maxwell equations due to the presence of imperfections of small diameter, Asymptot. Anal., 30 (2002), 331--350.


\bibitem{WG0} S. Arnold, M. Khoshsima, I. Teraoka,
S. Holler, and  F. Vollmer, Shift of whispering-gallery modes in microspheres by
protein adsorption, Optics Lett., 28 (2003), 272--274.


\bibitem{berljafa2014rational} M. Berljafa, and S. Güttel, A Rational Krylov Toolbox for MATLAB, MIMS EPrint 2014.56, The University of Manchester, UK, 2014.

\bibitem{bethe} H.A. Bethe and J. Schwinger, \textsl{Perturbation Theory for
Cavities}, Massachusetts Institute of Technology, Radiation
Laboratory, Cambridge, 1943.

\bibitem{brakhage1965dirichletsche} H. Brakhage, and P. Werner, {\"U}ber das dirichletsche Aussenraumproblem f{\"u}r die Helmholtzsche Schwingungsgleichung, Archiv der Mathematik, 16 (1965), 325--329.

\bibitem{myself1} A. Dabrowski, Explicit terms in the small volume expansion of the shift of {N}eumann {L}aplacian eigenvalues due to a grounded inclusion
              in two dimensions, {J. Math. Anal. Appl.}, {456} (2017), {731--744}.

\bibitem{myself4} A. Dabrowski, {On the behaviour of repeated eigenvalues of singularly perturbed elliptic operators}, preprint, 2018.

\bibitem{WG1} M.R. Foreman, J.D. Swaim, and F. Vollmer, Whispering gallery mode sensors,
Advances in Optics and Photonics, 7 (2015), 168--240.

\bibitem{gop} J. Gopalakrishnan, S. Moskow, and F. Santosa, Asymptotic and numerical techniques for resonances of thin photonic
structures, SIAM J. Appl. Math.,  69 (2008), 37--63.

\bibitem{haroche1}  S. Haroche and J.-M. Raimond,  \textsl{Exploring the quantum. Atoms, cavities and photons}, Oxford Graduate Texts, Oxford University Press, Oxford, 2006. 

\bibitem{heider} P. Heider, Computation of scattering resonances for dielectric resonators, Comput. Math. Appl., 60 (2010), 1620--1632.

\bibitem{HeiderBerebicezKohnWeinstein} P. Heider, D. Berebichez, R.V. Kohn, and M.I. Weinstein, Optimization of scattering resonances, Struct. Multidiscip. Optim., {36} (2008), 443--456.

\bibitem{santosa} C.-Y. Kao and F. Santosa,
Maximization of the quality factor of an optical resonator,
Wave Motion, 45 (2008), 412--427.

\bibitem{kim} S. Kim and J.E. Pasciak, The computation of resonances in open systems using a perfectly matched layer, Math. Comput., 
78 (2009), 1375--1398.

\bibitem{haroche2} J.C. Knight, N. Dubreuil, V. Sandoghdar, J. Hare, V. Lef\`evre-Seguin, J.M. Raimond, and S. Haroche, Mapping whispering-gallery modes in microspheres with a near-field probe, Optics Lett., 20 (1995), 1515--1517.

\bibitem{haroche3} J.C. Knight, N. Dubreuil, V. Sandoghdar, J. Hare, V. Lef\`evre-Seguin, J.M. Raimond, and S. Haroche, Characterizing whispering-gallery modes in microspheres by direct observation of the optical standing-wave pattern in the near field, Optics Lett.,   21 (1996), 698--700. 

\bibitem{lam} C.C. Lam, P.T. Leung, and K. Young, Explicit asymptotic formulas for the positions, widths, and strengths of resonances in Mie scattering, J. Opt. Soc. Am. B, 9 (1992), 1585--1592.


\bibitem{lin1} J. Lin and F. Santosa, Resonances of a finite one-dimensional photonic crystal with a defect, SIAM J. Appl. Math., 73 (2013),
1002--1019.

\bibitem{lin2} J. Lin and F. Santosa, Scattering resonances for a two-dimensional potential well with a thick barrier, SIAM J. Math.
Anal.,  47 (2015), 1458--1488.


\bibitem{ludwig} D. Ludwig, Geometrical theory for surface waves, SIAM Rev., 17 (1975), 1--15.

\bibitem{martin2006multiple} P.A. Martin, \textsl{Multiple scattering: interaction of time-harmonic waves with N obstacles}, 
Encyclopedia of Mathematics and its Applications, 107, Cambridge University Press, Cambridge, 2006. 

\bibitem{matkowsky} B.J. Matkowsky, A boundary layer approach to the whispering gallery phenomenon, Quart. Appl. Math., to appear.

\bibitem{agallery1} B.  Min, E. Ostby, V. Sorger, E. Ulin-Avila, L. Yang, X. Zhang, and  K. Vahala,
High-Q surface-plasmon-polariton whispering-gallery microcavity, Nature, 457 (2009), 455--459.

\bibitem{agallery2} B.-T. Nguyen and D.S. Grebenkov, Localization of Laplacian eigenfunctions in circular, spherical, and elliptical domains, SIAM J. Appl. Math., 73 (2013), 780--803. 


\bibitem{GW2} A.N. Oraevsky, Whispering-gallery waves, Quantum Electronics, 32 (2002), 377--400.

\bibitem{osting} B. Osting and M. Weinstein, Long-lived scattering resonances and Bragg structures, SIAM J. Appl. Math., 73 (2013), 827--852.

\bibitem{rapun2006indirect} M.L. Rap{\'u}n and F.J. Sayas, Indirect methods with Brakhage-Werner potentials for Helmholtz transmission problems, Numerical Mathematics and Advanced Applications, Springer, Berlin, Heidelberg, (2006) 1146--1154.


 \bibitem{GW3} G.C. Righini, Y. Dumeige, P. F\'eron, M. Ferrari, G. Nunzi Conti,
D. Ristic, and S. Soria, Whispering gallery mode microresonators: Fundamentals
and applications, Rivista Del Nuovo Cimento, 34 (2001), 435--488.

\bibitem{koenderink1} F. Ruesink, H.M. Doeleman, R. Hendrikx, A.F. Koenderink, and E. Verhagen, Perturbing open cavities: Anomalous resonance frequency shifts
in a hybrid cavity-nanoantenna system, Phys. Rev. Lett., 115 (2015), 203904.


\bibitem{steinbach2017combined} O. Steinbach  and U. Gerhard, Combined boundary integral equations for acoustic scattering resonance problems, Math. Meth. Appl. Sci., 40 (2017), 1516--1530.


\bibitem{WG4} F. Vollmer and S. Arnold, Whispering-gallery-mode biosensing: label-free
detection down to single molecules,
Nature methods, 5 (2008), 591--596.

\bibitem{WG5} F. Vollmer, S. Arnold, and D. Keng,  Single virus detection from the reactive shift of a whispering-gallery mode, Proc. Natl. Acad. Sci. USA,
105 (2008), 20701--20704.

\bibitem{majda} M. Wei, G. Majda, and W. Strauss, Numerical computation of the scattering frequencies for acoustic wave equations,
J. Comput. Phys.,  75 (1988), 345--358.




\end{thebibliography}
